# Enhanced spin-orbit coupling and charge carrier density suppression in $LaAl_{1-x}Cr_xO_3/SrTiO_3$ heterointerfaces

**Pramod Kumar, Anjana Dogra, P. P. S. Bhadauria, Anurag Gupta, K. K. Maurya and R. C. Budhani***

*National Physical Laboratory*, Council of Scientific and Industrial Research (CSIR)
*Dr. K.S. Krishnan Marg, New Delhi-110012, INDIA*

## Abstract

We report a gradual suppression of the two-dimensional electron gas (2DEG) at the $LaAlO_3/SrTiO_3$ interface on substitution of chromium at the Al sites. The sheet carrier density at the interface ($n_{\square}$) drops monotonically from ~ $2.2\times10^{14}$ $cm^{-2}$ to ~ $2.5\times10^{13}$ $cm^{-2}$ on replacing ≈ 60 % of Al sites by Cr and the sheet resistance ($R_{\square}$) exceeds the quantum limit for localization ($h/2e^2$) in the concentrating range 40 to 60 % of Cr. The samples with Cr ≤ 40 % show a distinct minimum ($T_m$) in metallic $R_{\square}(T)$ whose position shifts to higher temperatures on increasing the substitution. Distinct signatures of Rashba spin orbit interaction (SOI) induced magnetoresistance (MR) are seen in $R_{\square}$ measured in out of plane field ($H_{\perp}$) geometry at T ≤ 8K. Analysis of these data in the framework of Maekawa-Fukuyama theory allows extraction of the SOI critical field ($H_{SO}$) and time scale ($\tau_{SO}$) whose evolution with Cr concentration is similar as with the increasing negative gate voltage in LAO/STO interface. The MR in the temperature range 8K ≤ T ≤ $T_m$ is quadratic in the field with a +ve sign for $H_{\perp}$ and –ve sign for $H_{||}$. The behaviour of $H_{||}$ magnetoresistance is consistent with Kondo theory which in the present case is renormalized by the strong Rashba SOI at T < 8K.

*corresponding author: rcb@nplindia.org

# I. INTRODUCTION

Transition metal oxide heterostructures provide new opportunities to understand the effects of broken inversion symmetry on the interaction between spin, charge and orbital degrees of freedom in highly correlated electron systems [1]. One remarkable feature of epitaxial oxide heterostructures is the observation of a two dimensional electron gas (2*DEG*) at the interface of $LaAlO_3$ (LAO) and $LaTiO_3$ (LTO) epitaxial films grown on the $TiO_2$ terminated surface of (*001*) $SrTiO_3$ [2-5]. While $LaAlO_3$ is a band insulator of 5.2 eV energy gap, the properties of LTO are controlled by Mott physics due to the $3d^1$ configuration of $Ti^{3+}$ ions. However, in spite of these fundamental differences, the 2*DEG* at the LTO/STO and LAO/STO has broadly similar superconducting, magnetic and photoconducting properties [4-13]. As we move from Ti to V in the 3*d* transition metal series, the $LaVO_3$/STO interface also turns out to be metallic [14]. But this behaviour ceases to exist in $LaCrO_3$/STO and $LaMnO_3$/STO interfaces. However, while $LaCrO_3$ is an antiferromagnetic ($T_N$ = 290 K) band insulator, the manganite is a Mott insulator due to the $e_g^1$ electron. In both the cases, the interface turns out to be insulating with some magnetic character [15-17]. This dramatic change in the properties of the interface on going from $LaTiO_3$ to $LaMnO_3$ is yet to be understood, although one can attribute it to the increasing magnetic and mixed valent character of the higher 3*d* transition elements. In the context of doping, it is important to mention that the doping of the interface (δ-doping) of $LaAlO_3/SrTiO_3$ with few monolayers of $SrTi_{1-x}M_xO_3$ ($x$ = 0.02, and M a 3d transition metal) has been investigate and a quenching of the 2DEG behaviour is observed, presumably due to carrier localization at the 3d ion sites [18-19]. In the context of doping of the $LaAlO_3$ layer, recently Gray et. al. [20] have studied $LaAl_xR_{1-x}O_3$/STO system where the rare earth element R = Tm and Lu are only 2 % per formula unit. At such low doping level, the 2DEG remains more or less unaffected.

The other issue of importance in the LAO/STO problem is the co-existence of superconductivity and magnetic ordering. While the former has been established rigorously, the magnetic order is still a matter of debate due to its feeble nature [21-25]. The magnetic order manifests itself in subtle ways through the hysteresis in magnetoresistance and a characteristic minimum in resistivity. There is also a considerable debate on the source of carriers and magnetic moments whose condensation and ordering respectively leads to these two macroscopic phenomena. Generally, magnetism is attributed to $Ti^{3+}$ ions which are produced by oxygen vacancies. However, oxygen vacancies also contributes to carriers and it has been noticed that signature of magnetism are absent in high carrier density samples [26]. It is also important to notice that the carrier concentration in such a system is highly sensitive to growth conditions [27-29].

It is also established that the tetragonal distortion at the interface and confinement of electron gas in 2D lift the degeneracy of Ti ($d_{xy}$, $d_{yz}$ and $d_{xz}$) bands and indicates anomalous band dispersion, which can lead to a Lifshitz transition depending on band filling [30-33]. Since the distortion is intimately linked with the lattice constant of $LaAlO_3$, which is increased by Cr substitution, the latter is likely to affect the dispersion and occupancy of the $t_{2g}$ subbands.

To address the critical role of these transition elements in the electronic properties of the interface, we have introduced a controlled site substitution of Al by Cr in $LaAlO_3$ and have studied the electronic transport in $LaAl_xCr_{1-x}O_3/SrTiO_3$ interfaces. Here, on the one end of the composition phase space we have LAO/STO that shows the confinement of 2DEG at the interface, for the other end member $LaCrO_3/SrTiO_3$, no such confinement appears at the interface [15]. With the Cr substitution in $LaAl_xCr_{1-x}O_3/SrTiO_3$, one issue of interest is to look for the critical value of Cr substitution at which the interface becomes insulating. But more importantly, we would like to address the evolution of magnetotransport and collective phenomena such as magnetism and superconductivity and their bearing on the nature of the electron donating layer.

We note that the Cr substitution reduces the free carrier density at the interface in a rather gradual manner as samples with Cr concentration as high as 40 % remain metallic. The measurements of in-plane and out-of-plane field magnetoresistance in these systems shows enhanced Rashba spin-orbit interaction with the Cr addition. A comparison of our data with the results of electrostatic gating of LAO/STO interface suggests that the role of Cr addition is similar to that of the negative gate field.

## II. EXPERIMENT

Dense and compact laser ablation targets of $LaAl_{1-x}Cr_xO_3$ ($x$ = 0.0, 0.2, 0.4, 0.6, 0.8, 1.0), 25 mm in diameter and 6 mm thick were prepared by solid state reaction method. Films of thickness 20 unit cells (≈ 8nm) of all the compositions were deposited on the $TiO_2$ terminated $SrTiO_3$ (*001*) single crystal substrates by ablating these targets with a KrF excimer laser ($\lambda$ = 248 nm). The STO (*001*) substrates were pre-treated with the standard buffered hydrofluoric ($NH_4F$-HF) solution for 30 seconds in order to achieve a clean $TiO_2$ terminated surface [34]. Prior to the deposition, the chemically treated STO (*001*) were annealed for about an hour at 830 °C in the oxygen pressure of 7.4 x $10^{-2}$ mbar. Thin films of all the compositions were deposited at 800 °C with oxygen partial pressure of $1\times10^{-4}$ mbar. Deposition was performed by firing the laser at 1 *Hz* with an areal energy density of 0.56 to 0.65 $J.cm^{-2}$ on the targets. The film growth mode and thickness were monitored using RHEED (STAIB, 35 keV) during deposition. Once the deposition was completed, the samples were cooled down in the same $O_2$ partial pressure at the rate of 10 °C/min to ambient temperature.

For structural characterization, we have used X-ray diffraction and reflectivity measurements carried out in a PANalytical X'Pert MRD X-ray diffractometer equipped with a Cu $K_{\alpha 1}$ source. Magneto-transport measurements were performed in the Van der Pauw geometry using the resistivity attachment of a Quantum design magnetic property measurements system. Stable, low resistivity contacts were made with the 2DEG by wire bonding with thin Al wires.

## III. RESULTS AND DISCUSSION

A set of typical RHEED specular beam intensity oscillations observed as a function of growth time during the deposition of $x$ = 0 and $x$ = 0.6 films are shown in Fig. 1(*a* & *b*). Such intensity oscillations were used to control the thickness of the deposited films with a precision of one unit cell for all values of $x$. The observation of such sharp oscillations strongly indicates a Frank-van der Merve type [35] layer-by-layer growth of the $LaAl_{1-x}Cr_xO_3$ on STO. The RHEED patterns along the (*001*) crystallographic direction after film deposition are shown in insets of Fig. 1(*a* & *b*). Clear streaks, which are the characteristics of a 2*D* flat and crystallographically ordered surface, were preserved prior to during and after the growth.

The X-ray diffraction (XRD) patterns of ($x$ = 0.0, 0.2, 0.4 and 0.6) samples taken in $\theta$-$2\theta$ geometry are shown in Fig. 2(*a*). A distinct hump in diffracted intensity on the higher $2\theta$ side of the (*002*) reflection of the substrate marks the (*002*) Bragg line for the film. We have calculated the out-of-plane lattice parameter of the $LaAl_{1-x}Cr_xO_3$ films from the position of this hump and the similar ones observed next to the (*001*) and (*003*) reflections of the substrate, Variation of the out-of-plane *'c'* lattice parameter of the layers as a function of Cr concentration is shown in the inset of Fig. 2*a*. The increase in the *'c'* lattice parameter observed on Cr substitution in the film can be attributed to the larger ionic radius of $Cr^{3+}$ compared to that of $Al^{3+}$. The shift in the XRD peak of the film towards the reflection from the substrate with increasing Cr suggests that the top $TiO_2$ plane of STO is under less compressive strain because of a better lattice match between $SrTiO_3$ and $LaCrO_3$. The Phi scan around the (*022*) Miller plane exhibits four-fold symmetry with overlapping of the XRD peaks for the substrate and the film that indicates the cube-on-cube symmetry along (*011*) plane. Kiessig fringes were observed in XRR measurements (not shown here) which also confirm a layer-by-layer growth. Thus, the two complimentary measurement modes i.e. RHEED patterns along the (*001*) crystallographic direction and phi scan near (*022*) plane establish excellent crystalline quality of our films.

The temperature dependence of the sheet resistance $R_{\square}(T)$ of the $LaAl_{1-x}Cr_xO_3/SrTiO_3$ structure reveals a clear transition from a delocalized to localized electronic transport with the increasing $x$ as shown in Fig. 3. While the $R_{\square}(T)$ of the samples with $x$ = 0.0, 0.2, 0.4 and 0.6 decreases on lowering the temperature from 300 K to ≈ 50 K, the absolute value of $R_{\square}(T)$ at 300 K exceeds the quantum resistance for localization ($h/2e^2$) for the samples with $x > 0.4$. In $R_{\square}(T)$ of all the sample, a step like behaviour is noticed at ≈ 83 K and ≈ 183 K (see Fig. 3*a*) which also has been reported earlier in the heating cycle of resistance measurements [36-37]. The metallic behaviour seen in the samples with $x \leq 0.6$ gives way to a resistivity minimum followed by an increase in resistance on cooling below ≈ 50 K and finally a saturation at the lowest temperature. This behaviour is emphasized in Fig. 3(b) by plotting the data of metallic samples ($x$ = 0, 0.2 and 0.4) on an expanded scale and in Fig. 3 (c) by normalizing $R_{\square}(T)$ with respect to its value at 2 K. We note a shift of the resistivity minimum to higher temperature with the increasing Cr concentration while the magnitude of minimum is non-monotonic in these samples. A qualitative understanding of the $R_{\square}(T)$ can be made by considering various classical and quantum contributions to charge carrier scattering in a two dimensional metal. However, before we dwell upon detailed interpretation of the $R_{\square}(T)$ data, it is worthwhile to show how the $n_\square$ and $R_\square$ at 300 K vary as a function of Cr concentration. These data are shown in Fig. 4(b). We notice that for the Cr free sample the sheet carrier density $n_\square$ ($\approx 2.2 \times 10^{14}$ cm$^{-2}$) is close to the number predicted from the polarization catastrophe argument ($\approx 3.2 \times 10^{14}$ cm$^{-2}$) for the formation of 2*DEG* in LAO/STO. But it drops monotonically with a slope of $3.35 \times 10^{13}$ cm$^{-2}$ per atomic percentage of the Cr substitution. The Hall resistance ($R_H$) for all these samples measured at several temperatures over a field range of ±6 tesla remains linear in the field as shown in the Fig. 4*a*. In the simplest possible scenario, this suggest a single band electronic transport.

The temperature dependence of the sheet resistance of a complex system like this one will have contributions from processes over and above those related to classical scattering by frozen-in disorder, phonons and electrons. One important consequence of the diffusive motion of electrons in a two-dimensional metal at low temperatures is the enhanced quantum backscattering (*QBS*) as the inelastic scattering events diminish due to phonon freeze out. This effect leads to a *ln*T growth of $R_\square$ of the form,

$$\Delta R_\square(T) = - K(p/2)\ln[T/T_0] \qquad \ldots(1)$$

where $K$ is a constant and the index $p$ is related to the inelastic scattering time and $T_0$ is order of the Kondo temperature [38-39]. A similar temperature dependence of sheet resistance is also expected when quantum corrections to conductivity due to electron-electron interactions in a disordered two-dimensional conductor are taken into account [40]. While the *ln*T dependence of $R_\square$(T) seen in Fig. 3(c), over a limited window of temperature is suggestive of these models, the saturation of $R_\square$(T) at the lowest temperature is a deviation from Eq. 1. Also the temperature range over which the *ln*T dependence is seen is rather high for the *e-e* interaction to be operational.

In the presence of magnetic impurities, the classical *s-d* scattering [41] can contribute to resistance in such a manner that it diminishes quadratically in magnetization ($M$) in the concentrated impurity limit [42, 43]. At very low impurity concentration where the moments are not ordered, a Kondo-type electronic transport may prevail [44]. A careful examination of Hall resistance $R_H$ and magneto-resistance (*MR*) as a function of magnetic field does not reveal any hysteretic effects, suggesting the absence of direct magnetic scattering in the system down to 2 K. However a higher order *s-d* scattering in the Kondo sense can contribute to low temperature resistance in a characteristic manner if magnetic impurities are present in the dilute limit. The Kondo contribution leads to a resistance $\Delta R_K \sim B\,ln T$, where B is a function of exchange integral $J$ between electronic and impurity spins and the density of states at the Fermi energy [44]. At very low temperatures, the logarithmic divergence of $\Delta R_K$ is cut off due to the formation of Kondo singlets. We, therefore, model the R(T) data between 10 K to 100 K on the basis of the following generalized equation [43];

$$R_{\square} = R_0 + R_1T^2 + R_2T^5 + R_3\,ln T \qquad ...(2)$$

where $R_0$ is the residual resistance due to frozen-in disorder, the $T^2$ term comes from *e-e* scattering and the $T^5$ contribution has its origin in electron-phonon scattering. The last term of Eq.2 represents quantum correction to resistance, which may possibly come from weak localization, low temperature correction to *e-e* scattering and the scattering by local magnetic moments in the Kondo sense. It is clear from the fits shown in Fig. 3 (b) that Eq.2 expresses the temperature dependence of $R_{\square}$ well down to 10 K. However, below 10 K, we see the saturation of the resistance.

The other phenomenon which can significantly affect the low temperature electron transport in these systems is spin-orbit (*S-O*) interaction of the Rashba-type originating from the broken inversion symmetry. The conduction band in doped STO consists of $t_{2g}$ orbital, which under spin-orbit coupling splits into $J = 3/2$ and $J = 1/2$ states of energy separation ~ 17 meV [45-47]. In the context of the interfacial $t_{2g}$ orbitals in LAO/STO system, the quantum confinement along the *z* direction pushes the $d_{xy}$ orbitals down in energy compared to the $d_{zx}$ and $d_{yz}$. Thus $d_{xy}$ can be assigned to $J = 1/2$ atomic state where as $d_{zx}$ and $d_{yz}$ to $J = 3/2$. These energy bands are further split by the Rashba spin-orbit interaction, which is expected to be cubic in the wave vector *k* for $J = 3/2$ state. It has been shown that the *k*-linear and *k*-cubic Rashba splittings leave their distinctive signatures in the magnetoresistance (MR) [48]. The measurements of *MR* also permit us to differentiate between weak localization, quantum correction to the *e-e* scattering and the Kondo effect. While *WL* is an orbital effect and thus a strong negative magnetoresistance is predicted when the external field is perpendicular to the plane of the 2*D* gas ($H_\perp$ configuration), the *MR* due to *e-e* is small, positive and isotropic [40]. The Kondo contribution to *MR* is also small and isotropic but negative in sign.

In the presence of only *k*-cubic spin splitting as applicable in the case of $SrTiO_3$, the magneto-conductance correction due to *SOI* goes as [49];

$$\Delta[\sigma(B)-\sigma(0)] = \frac{e^2}{\pi h}\left\{\begin{array}{l}\psi\left(\frac{1}{2}+\frac{H_\varphi}{B}+\frac{H_{SO}}{B}\right)+\frac{1}{2}\psi\left(\frac{1}{2}+\frac{H_\varphi}{B}+\frac{2H_\varphi}{B}\right)-\frac{1}{2}\psi\left(\frac{1}{2}+\frac{H_\varphi}{B}\right)\\ -\ln\frac{H_\varphi+H_{SO}}{B}-\frac{1}{2}\ln\frac{H_\varphi+2H_{SO}}{B}+\frac{1}{2}\ln\frac{H_\varphi}{B}\end{array}\right\} \quad \ldots(3)$$

Here $H_\phi = h/8\pi eD\tau_\phi$ and $H_{SO} = h/8\pi eDe\tau_{SO}$, and $D$, $\tau_\phi$ and $\tau_{SO}$ are diffusion constant, dephasing time and spin-orbit scattering time respectively. The $\tau_{SO}$, in general, derives contribution from the scattering processes by phonons and impurities in the presence of spin-orbit interaction induced by heavy atoms in the lattice. This is called the Elliott-Yafet $\tau_{SO}$ [50,51]. However, the breaking of inversion symmetry at the interface of heterostructures gives rise to a stronger spin-orbit relaxation named the ‘Dyakonov-Perel’ (*DP*) mechanism [52]. In the present case we will assume that $\tau_{SO}$ is entirely dictated by the *DP* process.

We have performed two sets of *MR* measurements on the $LaAl_{1-x}Cr_xO_3$/STO samples whose sheet resistance is below the threshold for localization (≈12.9 kΩ). One, the transverse *MR* was measured with the magnetic field applied perpendicular to the plane ($H_\perp$) of the sample. Two, longitudinal MR was measured with the field applied parallel ($H_{||}$) to the plane of the *2DEG*. The *MR* is defined as $[(R_H - R_0)/R_0] \times 100\%$, where $R_H$ and $R_0$ are the sheet resistance in field and in zero-field condition respectively. The *MR* data for $H_\perp$ configuration in the field range of 0 to ±6 tesla at 2, 4, 6, 8, 10 and 15 K are shown in Fig. 5(*a-b*). It is worth mentioning here that widely varying magnitude and sign of MR have been reported in LAO/STO interfaces [53-56]. A comparative study of MR in film grown at high (~$10^{-4}$ mbar) and low (~ $10^{-6}$ mbar) $O_2$ partial pressure suggests that high MR is associated with samples where a significant fraction of carriers is derived from $O_2$ vacancies [57]. The very small values of MR and its sign in $H_\perp$ and $H_{||}$ geometrics are consistent with results on well controlled 2DEG samples [58, 59]. A careful examination of the four panels in Fig. 5 reveals that there are three distinct field dependence of the *MR* depending on the temperature of measurement. At the lowest T (= 2 K), the *MR* increases sharply with field, reaches a peak value and then drops on further increase in the field. In the ±*H* plots of Fig.5, the width of the *MR* cusp and the field at which it reaches a peak value increase with temperature and Cr concentration in the $LaAl_{1-x}Cr_xO_3$/STO films. Once the temperature exceeds a critical value $T_1$, the positive *MR* becomes nearly linear in the field. This H-linear behaviour gives way to parabolic field dependence once the temperature exceeds a critical value $T_2$. The values of $T_1$ and $T_2$ for $x$ = 0 and 0.4 samples are (6 K, 8 K) and (8 K, 10 K) respectively. The transition from linear to parabolic dependence is however gradual we also note that the peak magnitude of *MR* at 2 K increases with *x*.

We first discuss the +ve *MR* at $T>T_2$ which goes as $\sim H^{\alpha}$ ($\alpha > 1$). The +ve sign of the *MR* seen here is inconsistent with the *WL* and Kondo scattering pictures, and the $T_2$ is too large for quantum contributions of *e-e* to *MR* to be effective. This regime of *MR* can be attributed to the classical elastic and inelastic scattering events as the charge carriers tend to go in a cyclotron orbit in the out-of-plane field. The Kohler rule [59, 60] predicts, $\Delta R/R_0 \propto a\ |H/R_0|^2$. Its applicability however depends on the validity of the classical Boltzmann transport in this 2DEG system. Based on the Hall data and the fact that $\omega_c\tau << 1$, we believe the Kohlar rule can be applied to $H_\perp$ geometry MR. In the inset of Fig. 5 (*a',b'*) we have plotted the average *MR* ( = ½ (*MR*(+*H*) + *MR* (-*H*)) at 15 K for $x = 0$ and 0.4 samples as a function of $(H/R_0)^2$. For $H \geq 0.5$ Tesla we notice an approximate quadratic field dependence which is pronounced at the higher fields. From this MR data, the calculated Kohler mobility is ≈ 0.30 and 0.07 $cm^2V^{-1}s^{-1}$ for $x = 0.0$ and $x = 0.4$ samples respectively.

The data at $T \leq 6$ K show a clear deviation from the classical orbital effect. We interpreted these results in the framework of the *k*-cubic Rashba spin-orbit interaction effect which leads to a conductivity correction of the type expressed in Eq. 3. Fig. 6 (*a* and *b*) shows the fits to Eq. 3 in the field range of ±6 tesla. The quality of these fits is excellent. The critical fields, time scales and length scales extracted from such analysis are listed in the Table 1. For the intermediate temperature $T_1<T<T_2$, the data indicate a linear temperature dependence of MR. This is an interesting result in itself as linear MR has been seen in doped and disordered semiconductors and also in some topological insulators [61-65]. A tentative explanation for such dependence is the possibility of spatial fluctuations in carrier concentration and carrier mobility which can distort the carrier path as seen earlier in epitaxial layers of InAs and InAsN [66]. Such inhomogeneities can arise from the two different sources of carriers in LAO/STO and LTO/STO interface [67], in addition, since the three temperature regimes are closely spaced, one has to consider the effect of competition between the contributions of *S-O* and the classical orbital scattering. A rigorous analysis of the data in this temperature window should allow separation of these two contributions.

We see a striking difference in the field dependence of *MR* when the magnetic field is aligned parallel to the plane of the film. The results of these measurements for $x = 0.2$ and 0.4 samples performed at 2, 6, 10 and 20 K are shown in Fig. 7(a) and 7(b) respectively. While the *MR* curve at 2 K for $H_{||}$ has the same shape as in the case of $H_{\perp}$, the overall value of *MR* is smaller by a factor of two. There is a remarkable change in the sign and shape of the *MR* curve for T > 8 K. Now it appears quadratic in field, but with a negative sign which is not consistent with the Kohler type *MR* emanating from the classical orbital effect. We attributed it to the Kondo scattering where the magnetic field acts in the same manner as the temperature, the identity between the Kondo temperature ($T_K$) and Kondo field ($H_K$) goes as [65].

$$Sg\mu_B H_K = k_B T_K \qquad \ldots(4)$$

where $S$, $g$, $\mu_B$ and $k_B$ are impurity spin quantum number, Lande '*g*' factor, Bohr magnetron and Boltzmann constant respectively. The magnetic field dependence of sheet resistance in the Kondo regime can be modelled as [39,69,70].

$$R_{\square}(H_{//}) = R_{\square}(0) + R_k\,(H_{//}/H_k) \qquad \ldots(5)$$

To extract $H_k$, a magnetic field scale, related to Kondo temperature and g factor of the impurity spin [69], we apply Eq. 5 to the $H_{||}$ field data shown in Fig. 7(*a* and *b*) for 10 K. The best fitting obtained is $H_k = 80.5$ T and 85.5 T for $x = 0.20$ and 0.40 respectively.

A generalised expression of the type Eq. 5 can also be written for the Kondo contribution to the temperature dependence of zero field resistivity $R_K$ ($T/T_K$). Following Goldhaber Gorden [69] we have;

$$R_K\left(T/T_K\right) = R_K(T=0)\left[\frac{T_K^{'\,2}}{T^2 + T'^2_K}\right]^s \qquad \ldots(6)$$

Where $$T_K^{'} = \frac{T_K}{(2^{1/s} - 1)^{1/2}}$$

assuming only the Kondo contribution to resistance Eq. 6 dominated, a fit to the data shown in Fig. 3(d) yield a $T_K$ of ≈ 47.6 K and ≈ 58.5 K for $x = 0$ and $x = 0.40$ respectively. The $T_K$ and $H_K$ thus derived are in order of magnitude agreement to Eq. 4.

At this stage, we revisit the data given in Table 1 and compare the critical fields and length scales for different samples deduced in the frame work of Rashba SO contribution to magnetoresistance. *Caviglia et. al.* [72] has studied the behaviour of $H_\varphi$, $H_{SO}$, $L_\varphi$, $\tau_\varphi$ and $\tau_{SO}$ as a function of gate voltage in LAO/STO interface. The ratio of $H_{SO}/H_\varphi$ in their case is ≈ 50 at 1.5 K and zero bias, and increases with the increase in carrier concentration. For our undoped samples this ratio is ≈ 3 at 2 K, and decreases further with the Cr doping. This behaviour is consistent with the results of *Caviglia et. al.* as they also see a drop with the decreasing carrier concentration (increase negative gate voltage). The trend of $\tau_\varphi$ and $\tau_{SO}$ with increasing Cr concentration is also consistent with the effect of gating, although absolute numbers differ. It is worthwhile to compare our results with the data on 2D electron gas produced at the top gated (*001*) $SrTiO_3$ surface [73]. The absolute values of critical fields in this case are lower by an order of magnitude but their dependence on carrier concentration (gate field) is similar. It needs to be pointed out here that the carrier density is gated (*001*) STO is lower by a factor of ≈ 15 as compared to the sheet carrier concentration at the LAO/STO interface. Much lower critical fields have been reported in high mobility 2D gas in $Ga_xIn_{1-x}As$ quantum well structures [71].

## IV. CONCLUSIONS

In summary, we have successfully grown epitaxial $LaAl_{1-x}Cr_xO_3$ films on $TiO_2$ terminated $SrTiO_3$. The sheet resistance in the temperature range 2 K to 300 K increased significantly with substitution of Cr at the Al sites. For $x \geq 0.6$, the sheet resistance goes beyond the 2D quantum limit ($R_\square \approx h/2e^2$). The sheet carrier density is also suppressed monotonically with the increasing Cr substitution. The magnetoresistance data at $T \leq 20$ K have been analyzed in the detail to extract various classical and quantum contributions to electronic transport. $T \leq 10$ K out-of-plane field MR suggests strong Rashba spin-orbit coupling dominated transport. The in plane field MR and characteristic minimum in $R_\square(T)$ suggest Kondo-like scattering of charge carrier by localised spins associated with $Ti^{3+}$ and $Cr^{3+}$ ions. The effect of Cr substitution on magneto-transport characteristics mimics the behaviour of LAO/STO interface on negative gate bias, which reduces the carrier concentration at the interface.

## ACKNOWLEDGEMENTS

We acknowledge CSIR-India for financial support through AQuaRIUS Project. AD and RCB also acknowledge funding from the Indo-French Centre for Promotion Advanced Research. (Project # IFCPAR 4704-I), and PK thanks CSIR for a senior research fellowship.

## Table 1

| **Sample $LaAl_{1-x}Cr_xO_3$/STO** | **Tempera ture** | **$H_\varphi$ (Tesla)** | **$H_{SO}$ spin-orbit field (Tesla)** | **$L_\varphi$ (nm) Phase coherence length** | **D $=v_F^2\tau/2$ ($m^2$/sec) $\times 10^{-4}$** | **$\tau_\varphi$ (sec) $\times 10^{-12}$ de-phasing time** | **$\tau_{SO}$ (sec) $\times$ $10^{-13}$ spin-orbit scattering time** |
|---|---|---|---|---|---|---|---|
| *x = 0.0* | 2K | 0.17 | 0.72 | 31.25 | 8.22 | 1.18 | 2.79 |
| | 4K | 0.30 | 1.79 | 23.33 | 8.38 | 0.66 | 1.10 |
| | 6K | 0.43 | 2.08 | 19.60 | 8.64 | 0.44 | 0.92 |
| *x = 0.40* | 2K | 0.41 | 1.06 | 20.15 | 1.95 | 2.06 | 8.00 |
| | 4K | 0.50 | 1.34 | 18.10 | 1.99 | 1.66 | 6.19 |
| | 6K | 0.51 | 1.88 | 17.91 | 2.05 | 1.58 | 4.28 |

Parameters extracted from fitting the magnetoconductance data of Fig. 6 to Eq. 3. $H_\phi$ and $H_{SO}$ are the characteristic magnetic field, $\tau_\phi$ and $\tau_{SO}$ the relevant relaxation times and D the diffusion coefficient.

**Figure caption**

**Figure 1:** (Color online) Typical RHEED intensity oscillations and RHEED images recorded along the (*001*) crystallographic direction during the fabrication of (a) $x$ = 0 and (b) $x$ = 0.6. The clean intensity oscillations reveal a layer-by-layer growth of the $LaAl_{1-x}Cr_xO_3$ film on (*001*) $SrTiO_3$.

**Figure 2:** (Color online) (*a*) HRXRD around (*002*) peak of single crystal $SrTiO_3$ shows the film peak marked as 'F' adjacent to substrate peak, revealing epitaxial growth along the (*001*) crystallographic direction of the substrate. Inset of Fig. 2(*a*) shows the variation of out-of-plane lattice constant of the thin film as a function of Cr concentration. Fig. 2 (*b*) shows Phi scans around the (*022*) plane of the substrate and thin film. The four peaks separated at 90º show a four-fold symmetry, revealing epitaxial growth along (*001*) crystallographic direction of the substrate.

**Figure 3:** (Color online) (*a*) Temperature dependence of the sheet resistance of $LaAl_{1-x}Cr_xO_3/SrTiO_3$ heterostructures for $x$ = 0.0, 0.2, 0.4 and 0.6 in the temperature range of 2 K to 300 K. Figure also shows a clear insulating behaviour for sample with $x$ > 0.6. (*b*) Shows the $R_{\square}$(T) data for $x$ = 0.0, 0.2 and 0.4 samples from 2 K to 100 K. The solid lines are fit to Eq.1 in the temperature range of 10K to 100K. (*c*) Shows $R_{\square}$(T) of the three samples normalized with respect to its value at 2K to emphasize the minimum in $R_{\square}$(T). (d) Shows fitting of sheet resistance of $x$ = 0.0 and 0.40, including detailed Kondo term (Eq. 2 and Eq. 6) in the temperature range 2 K to 100 K.

**Figure 4:** (Color online) (*a*) Hall voltage divided by the applied current I as a function of magnetic field measured at 10K. (*b*) Variation of sheet resistance ($R_{\square}$) at 300 K and carrier concentration ($n_{\square}$) at 300 K and 10 K as a function of Cr concentration in $LaAl_{1-x}Cr_xO_3/SrTiO_3$.

**Figure 5:** (Color online) Magnetoresistance at different temperatures in transverse magnetic field ($H_{\perp}$) for $LaAl_{1-x}Cr_xO_3/SrTiO_3$ samples (*a, a'*) $x$ = 0.0 and (*b, b'*) $x$ = 0.40. Inset of figure (*a', b'*) shows Kohler plot at 15 K.

**Figure 6:** (Color online) Variation of magneto-conductance [$\Delta\sigma_{\square}(B) = \sigma_{\square}(B) - \sigma_{\square}(0)$], $\sigma$ being the sheet conductance, measured in a magnetic field applied perpendicular to the $LaAl_{1-x}Cr_xO_3/SrTiO_3$ interface at different temperatures (*a*) $x$ = 0.0, (*b*) $x$ = 0.40. Black line is the fit to *SOI* theory Eq. 3.

**Figure 7:** (Color online) Magnetoresistance at different temperatures in longitudinal field ($H_{||}$) from 6Tesla to -6 Tesla field for (*a*) $x$ = 0.20 and (*b*) $x$ = 0.40. Inset of figure 7(*a*) shows *MR* for $LaAlO_3/SrTiO_3$ in longitudinal mode

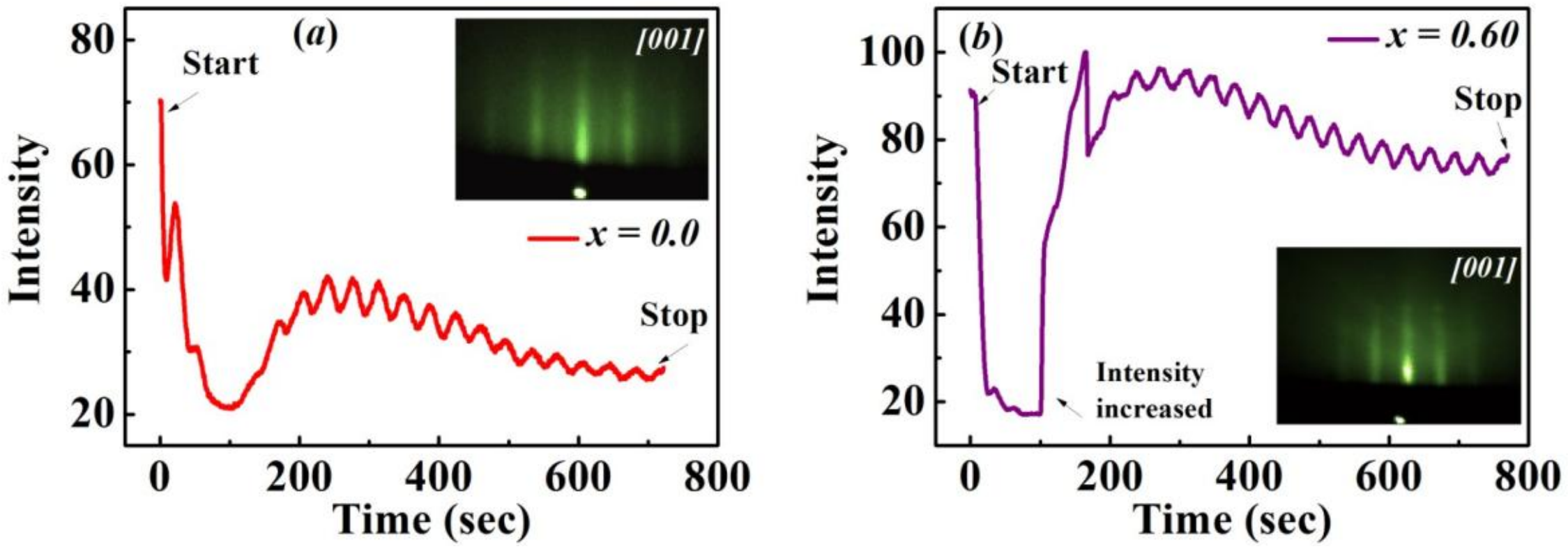

Figure 1

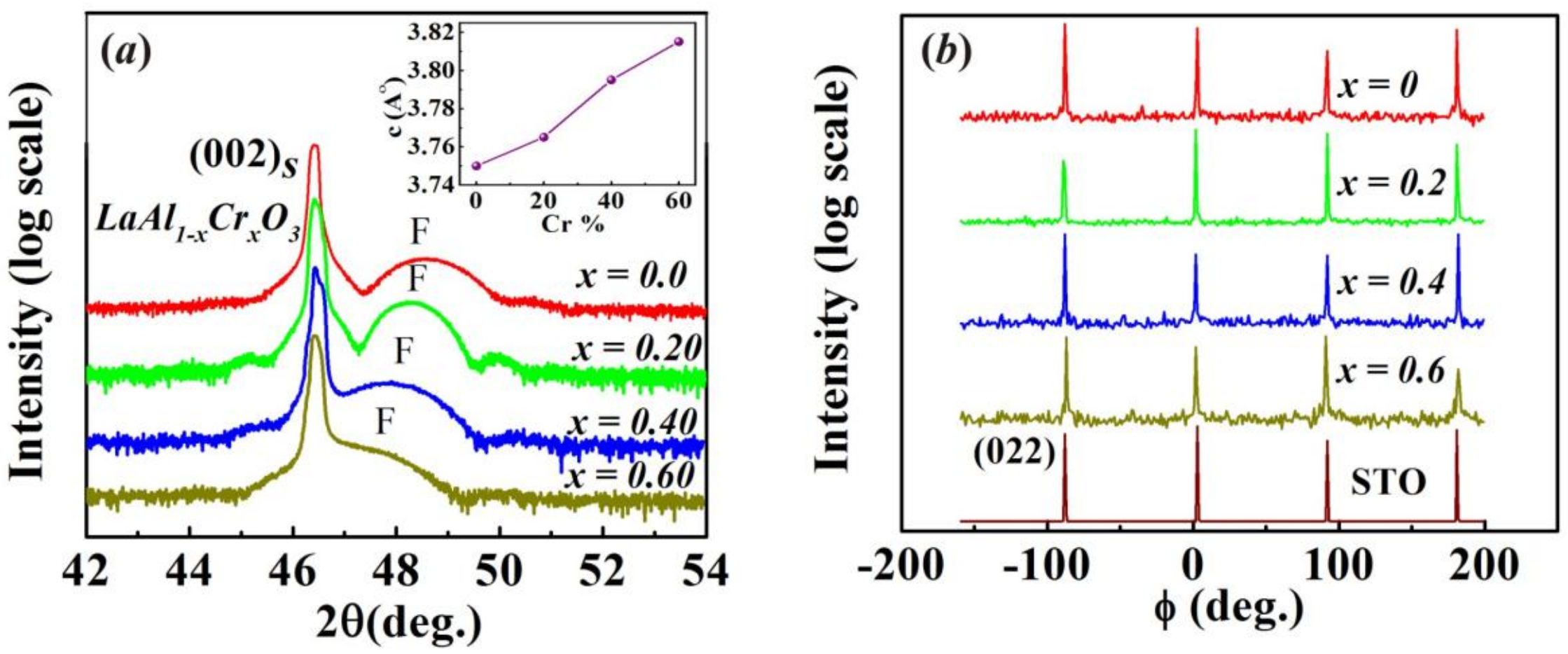

Figure 2

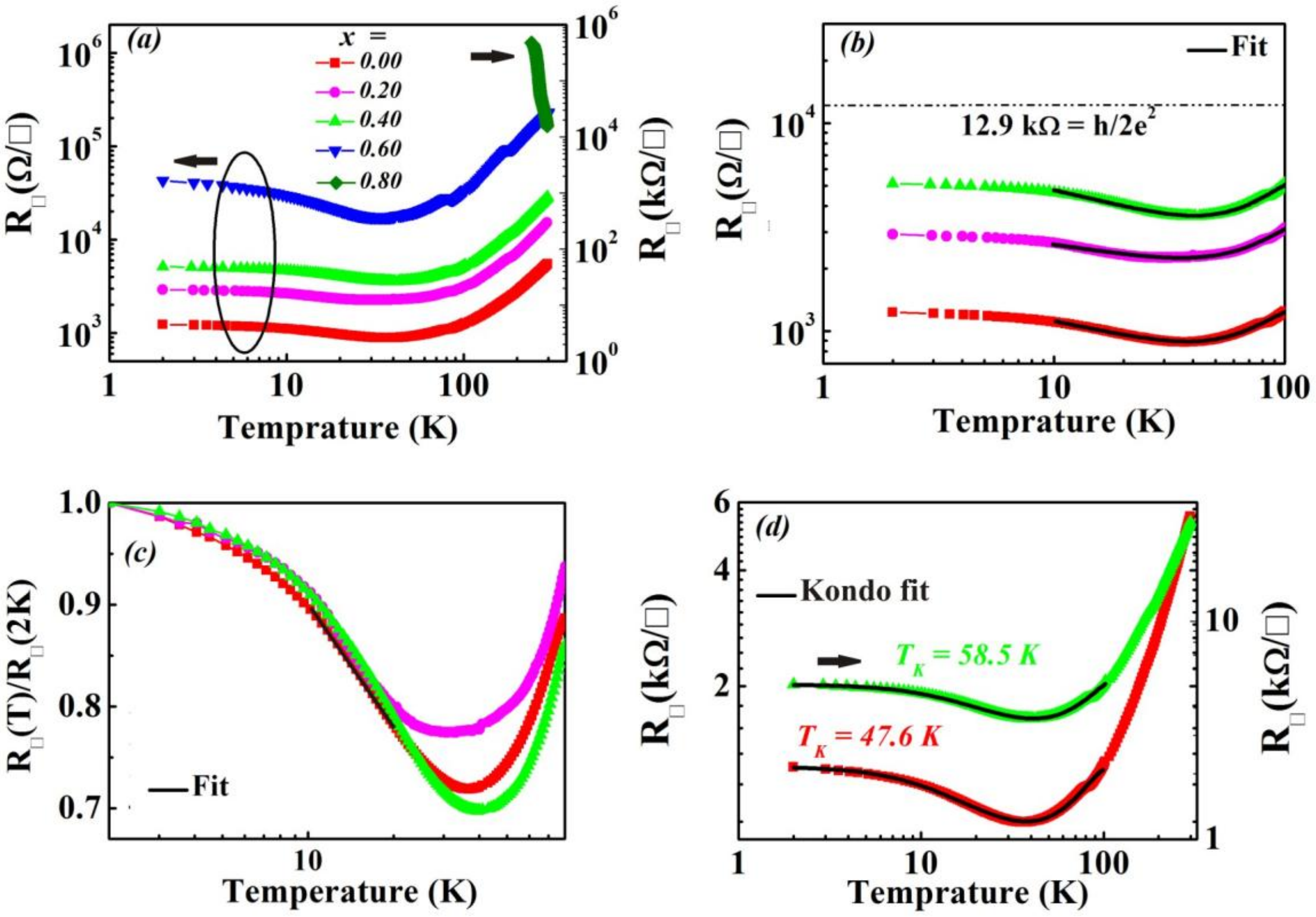

Figure 3

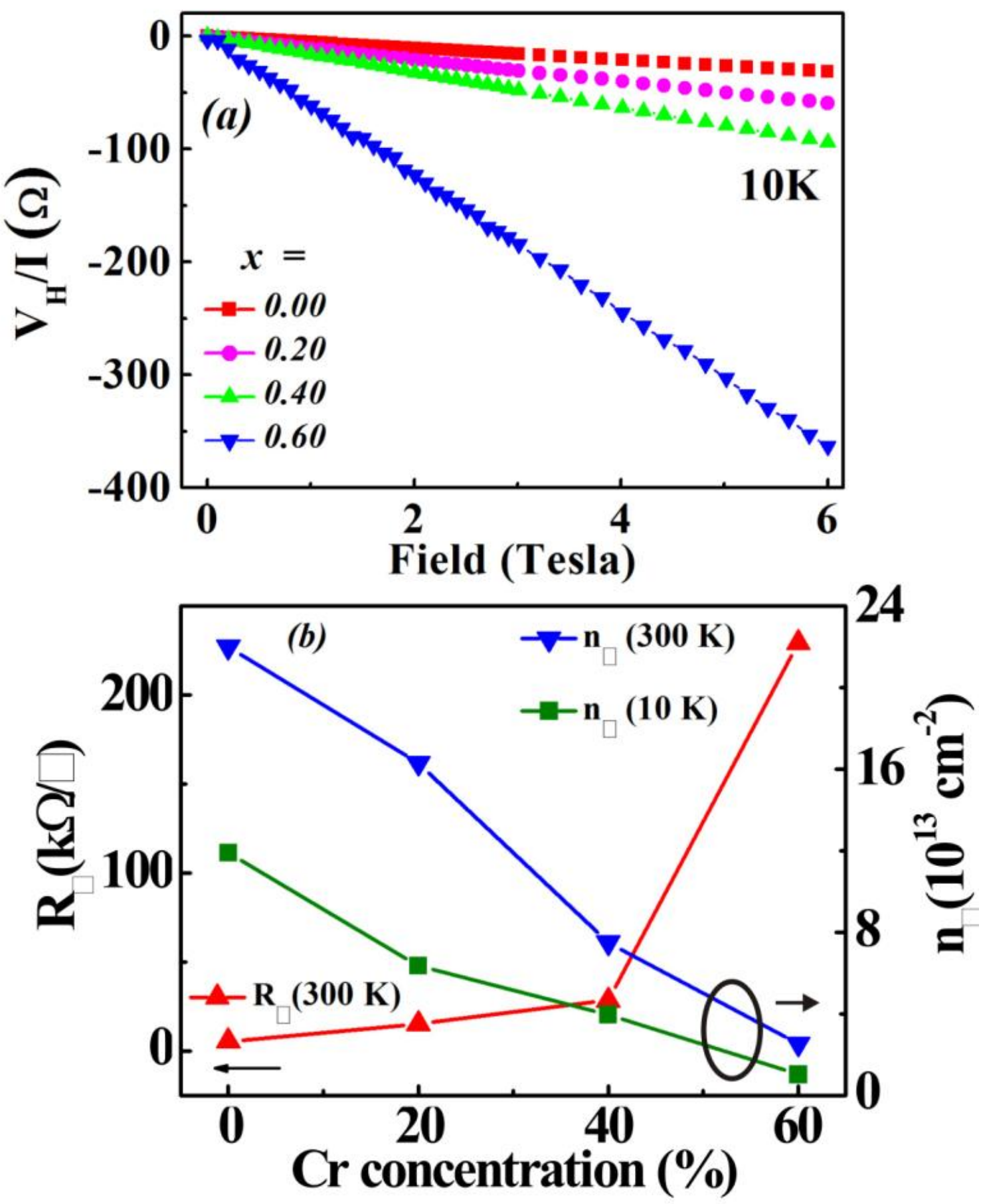

Figure 4

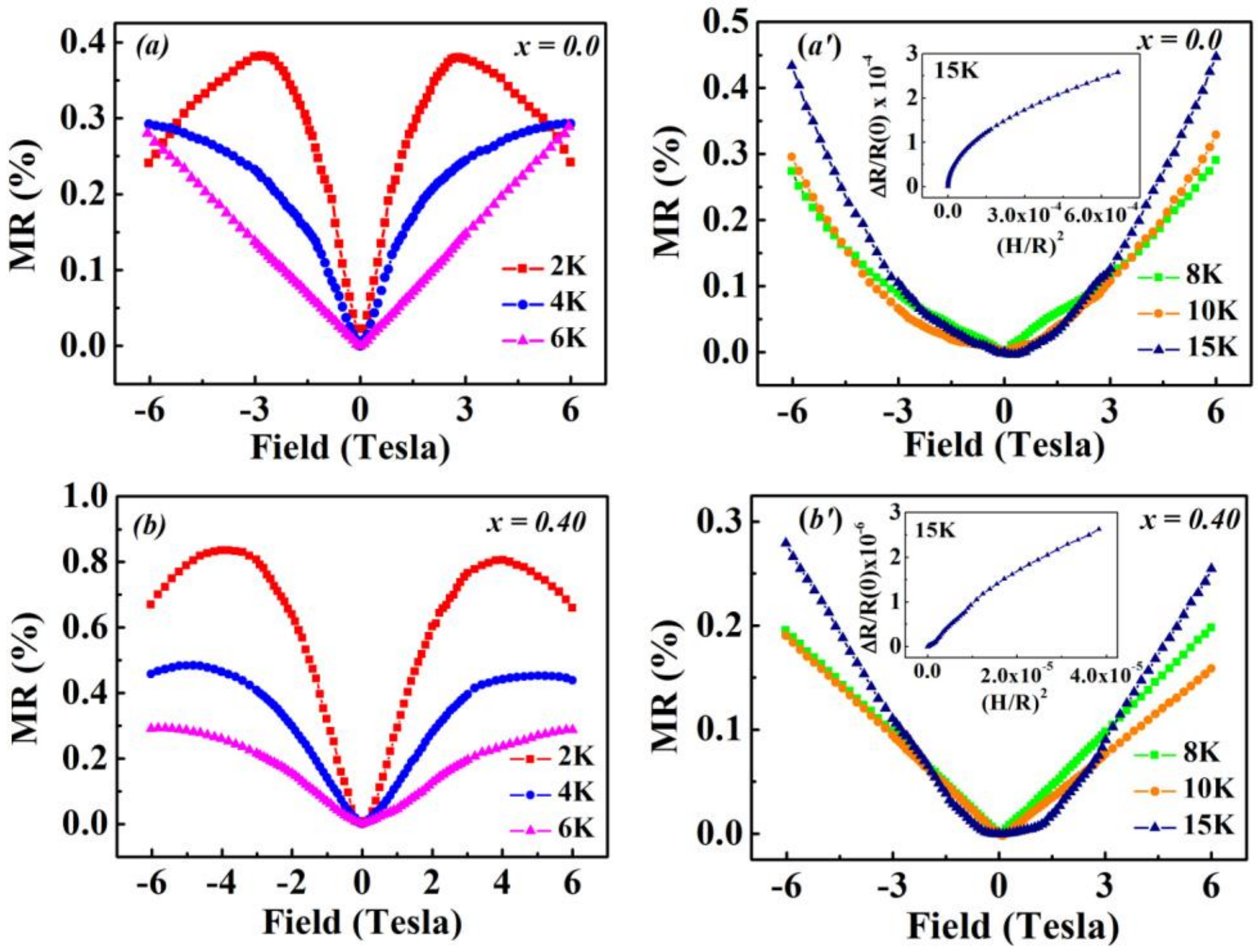

Figure 5

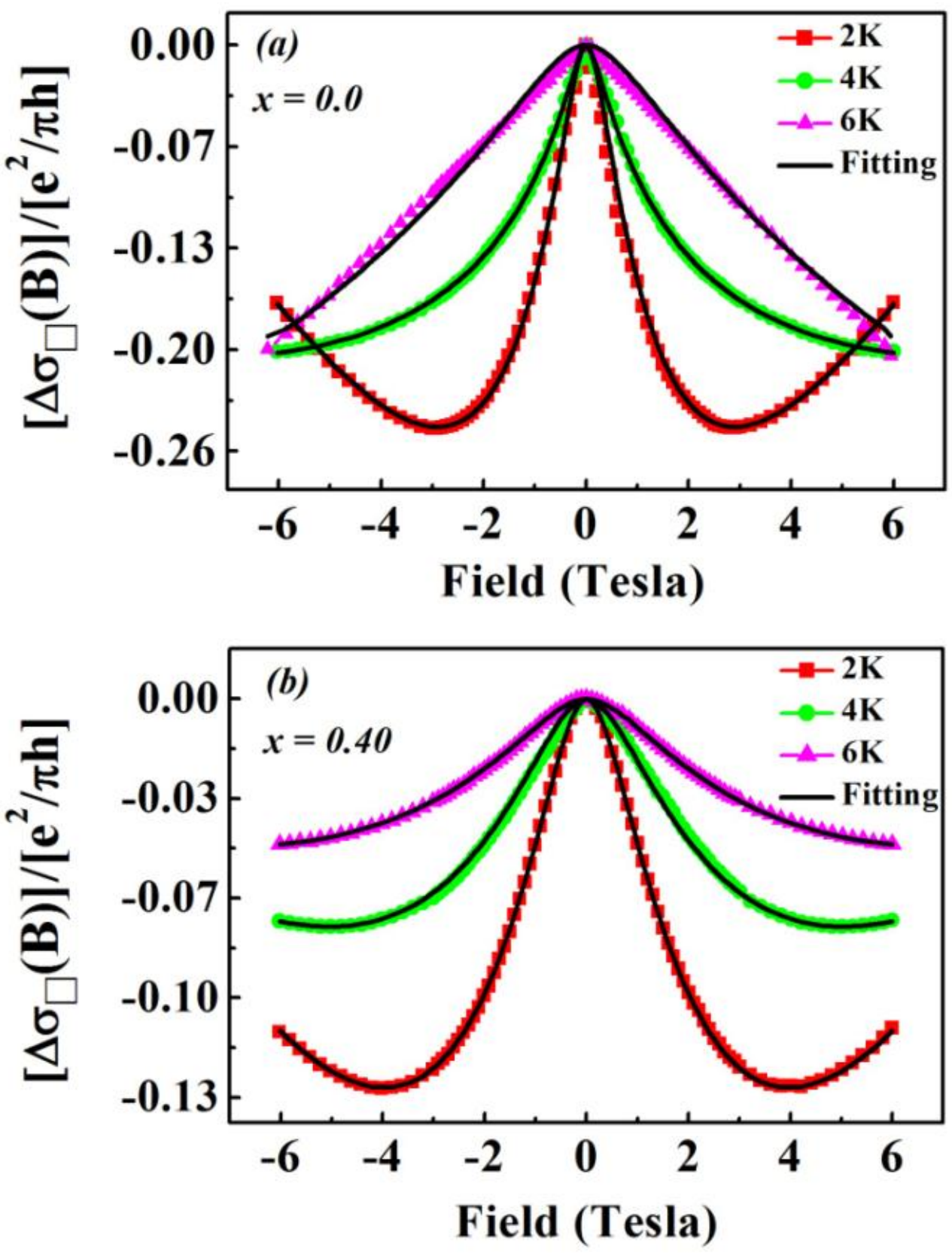

Figure 6

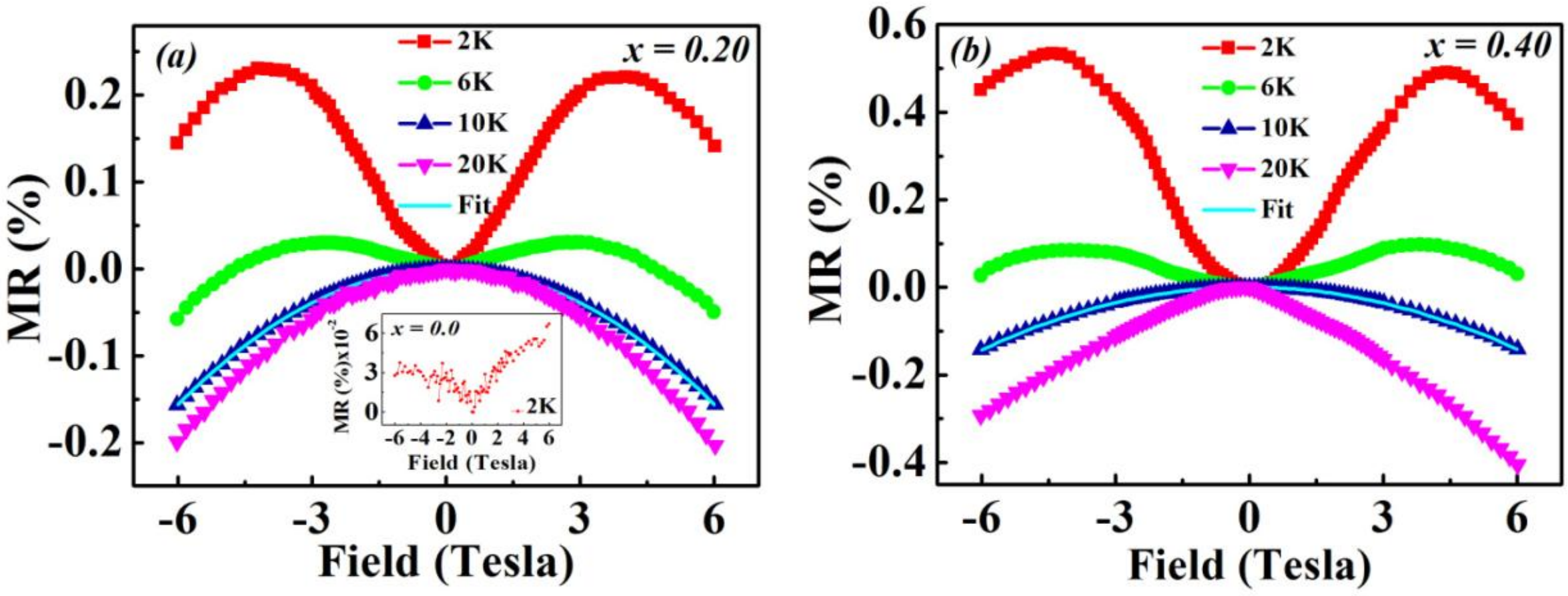

Figure 7